\documentclass[10pt, conference, letterpaper]{IEEEtran}

\usepackage{bbm}
\usepackage{amsmath}
\usepackage{amsthm}
\usepackage{amssymb}
\usepackage{verbatim}
\usepackage{graphicx}
\usepackage{subfigure}
\usepackage{url}
\usepackage{algorithm}
\usepackage{algorithmic}
\usepackage{color}
\theoremstyle{plain}
\newtheorem{lemma}{Lemma}

\newcommand{\argmax}{\operatornamewithlimits{argmax}}

\allowdisplaybreaks[4]

\newcommand{\prob}[1]{\mathbb{P}\left(#1\right)}

\newcommand{\indicator}{\boldsymbol{1}}

\newcommand{\floor}[1]{\left\lfloor #1 \right\rfloor}

\newcommand{\N}{\mathbb{N}}

\newboolean{showcomments}
\setboolean{showcomments}{true}
\newcommand{\jk}[1]{  \ifthenelse{\boolean{showcomments}}
{ \textcolor{red}{(JK says:  #1)}} {}  }
\newcommand{\todo}[1]{\ifthenelse{\boolean{showcomments}}
{ \textcolor{red}{(to do:  #1)}}{}}

\newcommand{\ignore}[1]{}

\begin{document}

% Set the width to be used for figures
\newlength{\figurewidth}\setlength{\figurewidth}{0.6\columnwidth}

\title{\fontsize{23}{23}\selectfont Adaptive flow-level scheduling for
  the IoT MAC}

\author{\IEEEauthorblockN{Pragya Sharma}
\IEEEauthorblockA{Department of Electrical Engineering\\
IIT Bombay}
\and
\IEEEauthorblockN{Jayakrishnan Nair}
\IEEEauthorblockA{Department of Electrical Engineering\\
IIT Bombay}
\and
\IEEEauthorblockN{Raman Singh}
\IEEEauthorblockA{Department of Electrical Engineering\\
IIT Bombay}
}

\maketitle

\begin{abstract}
Over the past decade, distributed CSMA, which forms the basis for
WiFi, has been deployed ubiquitously to provide seamless and
high-speed mobile internet access. However, distributed CSMA might not
be ideal for future IoT/M2M applications, where the density of
connected devices/sensors/controllers is expected to be orders of
magnitude higher than that in present wireless networks. In such
high-density networks, the overhead associated with completely
distributed MAC protocols will become a bottleneck. Moreover, IoT
communications are likely to have strict QoS requirements, for which
the `best-effort' scheduling by present WiFi networks may be
unsuitable. This calls for a clean-slate redesign of the wireless MAC
taking into account the requirements for future IoT/M2M networks. In
this paper, we propose a reservation-based (for minimal overhead)
wireless MAC designed specifically with IoT/M2M applications in
mind. The key features include: (i) flow-level, rather than packet
level contention to minimize overhead, (ii) deadline aware,
reservation based scheduling, and (iii) the ability to dynamically
adapt the MAC parameters with changing workload.
\end{abstract}

\section{Introduction}
\label{sec:intro}

%%Para 1
Over the past decade, WiFi has become the mainstay of non-cellular
wireless communication. It has been deployed widely across residential
as well as enterprise settings to provide seamless and high-speed
mobile internet access. It is estimated that over 94 million WiFi
hotspots were deployed worldwide as of 2016.\footnote{\tiny
  \url{https://www.worldwifiday.com/about-us/facts/}}

%%Para 2
WiFi is based on CSMA/CA (Carrier Sense Multiple Access/Collision
Avoidance) --- an entirely distributed medium access mechanism based
on channel sensing and collision avoidance using randomized backoff
\cite{Kurose2013}. The protocol operates at the link layer, providing
a best-effort delivery of packets from transmitter to receiver. In
line with the layered approach to networking, WiFi is oblivious to the
end-to-end flows that generate the packets it delivers, and is
therefore also blind to their Quality of Service (QoS)
requirements. However, WiFi works remarkably well in the settings in
which it is predominantly deployed: \emph{a moderate number of
  end-nodes requiring high data-rate connected to each access point}.

%%Para 3
However, several upcoming application scenarios differ considerably
from the settings in which WiFi is presently deployed. The explosion
of interest in the Internet of Things (IoT) and Machine to Machine
(M2M) communication points to scenarios where the density of connected
devices is projected to grow manifold in the coming
years.\footnote{\tiny
  \url{https://www.ericsson.com/en/mobility-report/internet-of-things-forecast}}
These IoT devices, which include household appliances, healthcare
devices, smart cars, sensors and actuators, will require reliable, but
not necessarily very high-speed internet access. In other words, in
contrast to current WiFi deployments, we should expect \emph{a
  considerable growth in the number of wireless end-devices, each of
  which will generate moderate, intermittent, but time-bound traffic.}
In such a setting, the overhead associated with the entirely
distributed and packet-level WiFi MAC is likely to become a
bottleneck. Moreover, this overhead, due to frequent collisions
between end-nodes attempting to access the channel, would also be
energy inefficient, which is a concern given that many IoT devices are
likely to be power constrained.

%%Para 4:
This paper proposes an alternative framework for MAC design,
particularly suited for upcoming IoT/M2M application settings: A large
number of wireless nodes connected to the internet via a single access
point, each generating moderate, occasional, but QoS sensitive
traffic. The proposed framework is based on Time Slotted Channel
Hopping (TSCH), which is supported under the IEEE 802.15.4e
specification \cite{De2016ieee}. The key features of the proposed
framework are:
\begin{enumerate}
\item The scheduling (specifically, admission control) is flow-aware,
  where a flow refers to a single burst of data generated by an IoT
  device.
\item Packet-level scheduling is performed centrally in an entirely
  reservation-based, QoS-aware manner, using ideas from the deadline
  scheduling literature for real-time systems.
\item Contention only takes place when the end-nodes attempt to
  register their flows with the access point. This reduces protocol
  overhead (relative to a MAC where contention takes place for the
  transmission of each packet).  Once a flow is admitted, it is
  centrally scheduled by the access point such that it meets its
  deadline.
\item The MAC parameters are dynamically adapted to the (possibly
  time-varying) traffic characteristics.
\item The framework supports highly heterogeneous end-devices, with
  widely ranging traffic patterns and energy constraints.
\end{enumerate}
As we demonstrate, the combination of flow-level, QoS-aware admission
control, and centralized reservation-based packet scheduling results
in a considerable gain in throughput as well as energy efficiency
relative to CSMA/CA.

It should be noted that while the primary intent of this paper is to
propose an alternative \emph{framework} for MAC design for IoT/M2M
applications, we describe an example \emph{protocol} based on this
framework with sufficient algorithmic and implementation detail to
enable a comparison with CSMA/CA.

The remainder of this paper is organised as follows. We describe the
setting for our MAC framework in Section~\ref{sec:setup}, the
admission control and scheduling aspects of the proposed framework in
Section~\ref{sec:adm_control_scheduling}, and the parameter adaptation
aspects in Section~\ref{sec:param_adaptation}. Finally, we evaluate
the performance of the proposed framework alongside CSMA/CA in
Section~\ref{sec:evaluation}.

\subsection*{State of the art}

Several papers have explored reservation-based wireless MAC
designs. The paper closest to the present one is \cite{EEDF}, which
also proposes a flow-level, deadline scheduling based MAC
design. However, in contrast with this work, \cite{EEDF} only
considers a single channel. Indeed, the possibility of scheduling over
multiple channels considerably complicates the deadline scheduling
problem \cite{Dertouzos1989}. Another related work
is~\cite{Octav2011}, which considers the problem of real-time
scheduling of periodic flows (as opposed to the `burst' flows
considered here) in a multi-hop, single channel setting.

The papers \cite{Zhu98} and \cite{Tinka2011} consider the problem of
decentralized slot reservation in a multi-hop environment. While
\cite{Zhu98} considers only a single channel, \cite{Tinka2011}
considers the multi-channel TSCH model as in the present
paper. However, both these papers consider packet-level reservation,
as opposed to the flow-level reservation considered here. Moreover,
neither of the above papers consider deadline-aware scheduling.

We are unaware of any work that considers the problem of deadline
constrained flow-level reservation in a multi-channel
setting. Moreover, none of the above papers consider online MAC
parameter adaptation, which is one of the key features of the present
work.

Finally, we note that several standards have been proposed for IoT/M2M
applications. While a comprehensive survey of these is beyond the
scope of the present paper due to space constraints, we note here that
802.11ah and 802.15.4e are among the prominent standards in this
space.~802.15.4e supports TSCH, upon which the proposed framework is
based.

%\vspace{-2mm}
%\clearpage
\section{Setting for MAC design}
\label{sec:setup}

In this section, we describe the setting for the proposed MAC design.

\subsection{Topology}

We consider a single-hop (star) topology, with a large number of IoT
nodes connected wirelessly to a central master node (a.k.a. hub,
access point). The master node has broadband access to the internet
(either wired or wireless), and is responsible for routing the traffic
generated by the IoT nodes to the intended destinations over the
internet.\footnote{For simplicity of exposition, we consider all
  traffic as \emph{upload} traffic from the IoT nodes. It is
  straightforward to see that our methodology extends to the general
  case where IoT nodes both transmit as well as receive data.} The IoT
nodes may be heterogeneous in the nature of traffic generated, their
QoS requirements, as well as their power constraints.

The above may capture a residential setting, where all IoT devices in
a household including appliances and healthcare sensors communicate
with a single access point, or an industrial setting, where all the
different sensors and actuators on a factory floor communicate with a
central controller.

\subsection{Traffic model}

The IoT nodes generate flows (transmission requests) sporadically,
which need to be transmitted to the master node. Each flow $i$ is
characterized by a load $l_i,$ which denotes the number of packets
that comprise the flow, and a deadline $d_i,$ which is the maximum
delay that can be incurred in transmitting all the $l_i$ packets to
the master node. A flow is considered to be successful if its load is
served (i.e., all its packets are transmitted to the master) within its
deadline.

The setting of interest is one where the number of IoT nodes connected
to the master is large, but each node generates flows only
occasionally. This is analogous to the case of telephone networks,
where a single switch (in the wireline setting) or base station (in
the wireless setting) serves a large number of subscribers, who
generate call requests occasionally. Thus, borrowing the modelling
framework from telephone networks, we assume that the generation of
flows (by all the IoT nodes connected to the master) follows a Poisson
process of rate $\lambda.$\footnote{Of course, the value of $\lambda$
  is unknown to the protocol or the master node. We describe our
  approach for optimally adapting the MAC parameters to the traffic
  characteristics in Section~\ref{sec:param_adaptation}.}  We note
that the assumption of Poisson arrivals is well backed by both
theoretical justification as well as empirical evidence
\cite[Chapter~11]{harchol2013performance}.

\subsection{Frame Structure}

\begin{figure}[t]
	\begin{center}
	\includegraphics[width=\columnwidth]{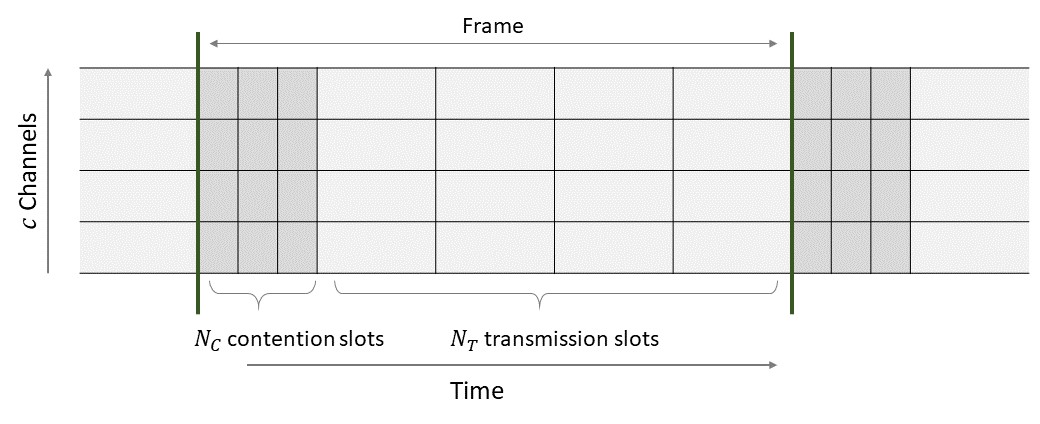}
	\caption{Structure of a frame in proposed MAC design}
	\label{fig:framework}
	\end{center}
\end{figure}

The proposed MAC design is based on Time Slotted Channel Hopping
(TSCH). Let $c$ denote the number of channels the protocol operates
on. These channels are assumed to be identical in capacity. It is
further assumed that an IoT node can transmit/receive on only a single
channel at a time, whereas the master node can receive on all $c$
channels simultaneously. Time is divided into frames, each frame
consisting of a contention phase (for new flows to get admitted with
the master) and a transmission phase (when the actual packet
transmissions take place); see Figure~\ref{fig:framework}.

Specifically, the contention phase consists of $N_C$ contention slots,
each one time unit long.\footnote{We will see that it is convenient to
describe time at the granularity of a contention slot.
%In practice, one time unit might be of the order of 10 $\mu$s, which
%is the order of the duration of a backoff slot under Wi-Fi.
} The transmission phase consists of $N_T$ transmission slots, each
$k$ time units long, where $k > 1$ is an integer. Each transmission
slot can support a single packet transmission. Thus, the length of the
frame equals $T = N_C + k N_T$ time units.\footnote{The master node
  would of course need to make regular broadcasts announcing the MAC
  parameters to be used by the IoT nodes, and the transmission
  schedules to be followed over each frame. For simplicity, we ignore
  the time spent for these broadcasts in our frame structure.}

Note that there are $c N_C$ contention blocks in each frame (a block
referring to a time slot on a particular channel); these are used by
newly arrived flows to register with the master, as described in
Section~\ref{sec:adm_control_scheduling}. The master performs
admission control, and schedules the accepted flows in the
transmission phase, taking into account the deadlines of the different
accepted flows (details in
Section~\ref{sec:adm_control_scheduling}). Note that there are $cN_T$
transmission blocks in each frame.\footnote{The values of $N_C$ and
  $N_T$ are themselves adapted by the protocol based on the traffic
  characteristics, as described in
  Section~\ref{sec:param_adaptation}.}

%\vspace{-2mm}
\section{Admission Control and Scheduling}
\label{sec:adm_control_scheduling}

In this section, we present the details of the proposed MAC design,
including the contention process for newly generated flows, admission
control, and scheduling of admitted flows. The adaptation of the MAC
parameters based on the observed traffic characteristics is described
in Section~\ref{sec:param_adaptation}.

\subsection{Contention}
\label{sec:contention}

As noted in Section~\ref{sec:setup}, we assume that flows are
generated by the IoT nodes according to a Poisson process of rate
$\lambda.$ Each generated flow $i$ is associated with the tuple
$(t_i,l_i,d_i),$ where $t_i$ denotes the generation time, $l_i$
denotes the load measured in number of packets (i.e., number of
transmission blocks required by the flow), and $d_i$ denotes the
deadline (i.e., the flow must complete all $l_i$ transmissions until
time $t_i + d_i$ in order to be considered successful).

The proposed contention mechanism works as follows. Each flow that is
generated over the duration of any frame has the chance to contend for
admission during the contention phase of the following
frame. Specifically, each flow contends for admission with probability
$p,$ where $p,$ which we refer to as the contention probability, is a
protocol parameter whose value is determined (and broadcast
periodically) by the master node. Each contending flow $i$ picks a
contention block (out of the $cN_C$ possibilities) uniformly at
random, and transmits an admission request in that block, which
contains all relevant flow information, including the tuple
$(t_i,l_i,d_i).$ If multiple contending nodes pick the same contention
block, their admission requests collide and are not received by the
master. On the other hand, if a certain contention block is selected by
exactly one contending flow, its admission request is received by the
master, and is included for consideration in the admission control
process (described in Section~\ref{sec:admission}).

The contention probability $p$ is set so as to maximize the number of
admission requests that are successfully received by the master. It is
instructive at this point to characterize the optimal value of $p.$
Note that the number of generated flows over a frame is
$\mathrm{Poisson}(\lambda T).$\footnote{Here,
  $\mathrm{Poisson}(\lambda)$ denotes a Poisson random variable with
  parameter $\lambda.$} Thus, the number of contending flows that
transmit in any particular contention block is
$\mathrm{Poisson}(\frac{\lambda Tp}{cN_C}).$ As a result, the
probability of a successful admission request from any particular
contention block equals $$\prob{\mathrm{Poisson}(\frac{\lambda
    Tp}{cN_C}) = 1} = \frac{\lambda Tp}{cN_C} e^{-\frac{\lambda
    Tp}{cN_C}}.$$ We conclude that the expected number of admission
requests received during one contention phase equals $$\lambda
Tpe^{-\frac{\lambda Tp}{cN_C}}.$$ It is now easy to see that the value
of $p$ that maximizes the expected number of successful admission
requests is given by
\begin{equation}
  \label{eq:cont_prob_opt}
  p^* = \min(1,\frac{c N_C}{\lambda T}).
\end{equation}

As expected, the optimal contention probability is inversely
proportional to $\lambda$ (for $\lambda > \frac{cN_C}{T}$). This is
analogous to the optimal transmission probability under slotted Aloha
being inversely proportional to the number of transmitting nodes
\cite{BG1992}.

Of course, since the master node does not know the value of~$\lambda,$
it cannot directly set the contention probability to its optimal
value. In Section~\ref{sec:param_adaptation}, we describe an iterative
mechanism for the master to \emph{learn} $p^*$ based on the observed
collision statistics.

\subsection{Admission Control}
\label{sec:admission}

We now describe the mechanism by which the master node selects which
admission requests to admit, based on the loads and deadlines
associated with the requests.

Given the (say $n$) admission requests received at the end of the
contention phase, the master constructs a list of these admission
requests as follows: $$\mathit{NewRequests} = ((l_1,\hat{d}_1),
(l_2,\hat{d}_2),\cdots, (l_n,\hat{d}_n))$$ Here, $l_i$ is the number
of transmission blocks requested by Flow~$i$, and $\hat{d}_i$ is the
deadline of the flow from the present time, also measured in number of
transmission slots. Specifically, $\hat{d}_i$ is the number of
transmission slots in the future by when Flow~$i$ needs to be
scheduled $l_i$ times in order to be successful.\footnote{The
  transformation of the deadline from time units to number of
  remaining transmission slots is straightforward. If $t$ denotes the
  time at the end of the contention phase in which a flow request $i$
  is received, its deadline in transmission slots is given
  by $$\hat{d}_i = N_T \floor{\frac{t_i + d_i - t}{T}} +
  \min\left(N_T, \floor{\{\frac{t_i + d_i - t}{T}\} \frac{T}{k}}
  \right),$$ where $\{x\} = x - \floor{x}$ denotes the fractional part
  of $x.$}

Additionally, the master maintains a list of (say $m$) \emph{active
  flows}, which have been previously admitted, but not yet
completed. This list is defined as follows:
$$\mathit{ActiveFlows} = ((l_1,\hat{d}_1),
(l_2,\hat{d}_2),\cdots, (l_m,\hat{d}_m))$$ In the above list, $l_i$
denotes the residual load of Flow~$i,$ i.e., the number of packets
remaining to be transmitted, and $\hat{d}_i$ denotes the remaining
deadline, i.e., the remaining number of future transmission slots in
which the flow needs to be completed.

Given these two lists, the master seeks to admit the largest number of
admission requests, given the residual service requirements of the
existing flows. It is well known that when $c > 1,$ it is impossible
to optimally admit and schedule the largest number of admission
requests in an online fashion \cite{Dertouzos1989}, so it is necessary
to employ a reasonable heuristic. The proposed algorithm for selecting
which of the admission requests to accept is the following.

\begin{algorithm}
  \caption{Admission Control}
  \begin{algorithmic}[1]
    \STATE 
    Sort $\mathit{NewRequests}$ in increasing order of $l_i$
    \STATE
    $n \gets \mathnormal{length}(\mathit{NewRequests})$ 
    \FOR{$i = 1$ to $i = n$}
    \STATE
    $\mathit{Flows} \gets \mathit{ActiveFlows}$
    \STATE 
    Append $\mathit{NewRequests}[i]$ to $\mathit{Flows}$    
    \IF{$\mathnormal{FeasibilityCheck}(\mathit{Flows})$}
    \STATE
    Append $\mathit{NewRequests}[i]$ to $\mathit{ActiveFlows}$    
    \ENDIF
    \ENDFOR
  \end{algorithmic}  \label{algo:adm_control}
\end{algorithm}
Note that the algorithm first sorts the new admission requests in
order of increasing load, and sequentially admits each admission
request in the list if it can be feasibly scheduled along with the
already admitted flows. The basis of this admission control algorithm
is a boolean function $\mathnormal{FeasibilityCheck}(S),$ which
returns \emph{true} if the set $S$ of flows can be feasibly scheduled,
i.e., if there exists a schedule that allows each flow in $S$ to be
completed before its deadline.

There are several ways of implementing the above feasibility check. One
is based on the classical Least Laxity First (LLF) scheduling
algorithm (see, for example, \cite{Dertouzos1989}). The laxity of a
flow $i$ is defined as the difference between the remaining deadline
and the remaining load, i.e., $\hat{d}_i - l_i.$ Note that laxity is
an indicator of the urgency of the flow; a flow with laxity zero
\emph{must} be scheduled in order to be successful. The Least Laxity
First algorithm, as the name suggests, schedules in each transmission
slot the $c$ flows with the least laxity (with ties broken
arbitrarily). If all flows complete before their deadline (i.e., the
laxity remains non-negative until completion) under LLF scheduling,
then the corresponding set of flows is deemed feasible.
\ignore{
\begin{algorithm}
  \caption{LLF Feasibility Check}
  \begin{algorithmic} [1]
    \STATE Input: $\mathit{Flows} = ((l_1,\hat{d}_1), (l_2,\hat{d}_2),\cdots,
(l_n,\hat{d}_n))$
    \FOR{$t = 1$ to $\max(\hat{d}_1,\hat{d}_2,\cdots,\hat{d}_n)$}
    \IF{$\mathit{Flows}$ is empty}
    \RETURN TRUE
    \ENDIF
    
    \STATE $S \gets$ set of $\min(c, \mathrm{length}(\mathit{Flows}))$
    elements of $\mathit{Flows}$ with least laxity
    
    \FOR{$(l_i,\hat{d}_i) \in S$}
    \STATE $l_i \gets l_i - 1$ 
    \STATE $\hat{d}_i \gets d_i - 1$
    \ENDFOR
    
    \FOR{$(l_i,\hat{d}_i) \notin S$}
    \STATE $\hat{d}_i \gets d_i - 1$
    \ENDFOR
    
    \FOR{$(l_i,\hat{d}_i) \in \mathit{Flows}$}
    \IF{$l_i == 0$}
    \STATE Remove $i$ from $\mathit{Flows}$
    \ENDIF
    \IF{$\hat{d}_i - l_i < 0$}
    \RETURN FALSE
    \ENDIF
    \ENDFOR
    
    \ENDFOR
    \RETURN TRUE
  \end{algorithmic}  \label{algo:llf}
\end{algorithm}
}
The correctness of this feasibility check is guaranteed by the
following result.
\begin{lemma}[\cite{Dertouzos1989}]
  If a given set of concurrent flows
  $\{(l_1,\hat{d}_1),(l_2,\hat{d}_2,\cdots \}$ can be successfully
  scheduled by any algorithm, then they can be scheduled successfully
  using LLF.
\end{lemma}

Another approach is to pose the feasibility check as a max flow
problem on an edge-capacitated directed acyclic graph (see Chapter~5
in \cite{Brucker2007}). We omit the details here.

\subsection{Scheduling}
\label{sec:scheduling}

Once the master node decides which of the admission requests to accept
(right after the contention phase), it schedules the active flows in
the present transmission phase. (Note that the admission control
process ensures that all the accepted flows can indeed be scheduled
before their deadlines.) This schedule is constructed by applying the
LLF algorithm for the $N_T$ transmission slots in the transmission
phase of the current frame.

%\vspace{-2mm}
\section{MAC parameter adaptation}
\label{sec:param_adaptation}

The MAC design described in Section~\ref{sec:adm_control_scheduling}
has two key parameters: (i) the contention probability $p,$ and (ii)
the fraction of time in each frame dedicated to the contention phase
(as determined by $N_C$ and $N_T$). Note that a larger contention
phase allows more flow requests to be received by the master, but
leaves less time for the actual data transmissions. These parameters
need to be optimized based on the (apriori unknown) statistics of the
traffic generated by the IoT nodes. In this section, we describe
online adaptation approaches for optimally setting both the above
parameters.

We begin with the contention probability adaptation.

\subsection{Estimating $p^*$ online}

Note that the optimal value of the contention probability $p^*$ is
given by~\eqref{eq:cont_prob_opt}. One approach to estimating $p^*$
would be to estimate the flow arrival rate $\lambda$ based on the
observed number of idle, collision, and successful contention blocks
during the contention phase of successive frame, and to set the
contention probability as a function of the estimated $\lambda.$ This
would be in line with the proposals to perform node cardinality
estimation in wireless networks to enable optimization of MAC
parameters \cite{Qian2011,Kadam2017}.

\begin{figure}[!hbt]
	\begin{center}
	\includegraphics[width = 0.4\textwidth]{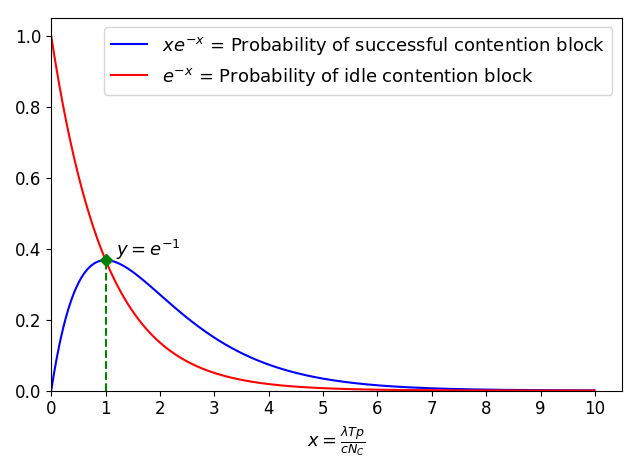}
	\caption{Probability of number of successful and idle blocks}
	\label{fig:graph}
	\end{center}
\end{figure}

However, we propose a simpler, direct method for estimating $p^*.$
Consider Fig.~\ref{fig:graph}, where we plot as a function of $x =
\frac{\lambda Tp}{cN_C},$ the probability that a contention block is
idle ($\prob{\mathrm{Poisson}(x) = 0}$), and the probability that a
contention block results in a successful admission request generation
($\prob{\mathrm{Poisson}(x) = 1}$). Note that the optimal choice of
$x$ equals $\min(1,\frac{\lambda T}{cN_C}).$ Thus, the optimal
contention probability corresponds to setting the probability of an
idle contention block as close to $1/e$ as possible, subject to the
constraint $p \in [0,1].$

The monotonicity of the probability of an idle block as a function of
$p$ then suggests a simple stochastic approximation scheme for
adapting $p.$ Let $N_I(t)$ denote the number of idle contention blocks
observed during the contention phase of frame $t.$ We adapt $p$ as
follows.
\begin{equation} \label{eq:p_est}
p_{t+1} = \left[ p_t + \delta_t \left(\frac{N_{I}(t)}{cN_C} -
  \frac{1}{e}\right) \right]_{[0,1]}
\end{equation}
Here $\delta_t$ is the step size, and $[x]_{[0,1]} =
\min(1,\max(x,0))$ denotes the projection of $x$ on the interval
$[0,1].$ Mathematically, the convergence of $p_t$ to $p^*$ can be
proved under a suitably diminishing step size sequence by standard
techniques \cite{Borkar2009SA}. However, to make the adaptation robust
to (slow) changes in the arrival rate, we take a fixed step size,
i.e., $\delta_t \equiv \delta.$

Finally, we note that the optimal contention probability depends
on the value of $N_C,$ which the preceding presentation assumes is a
constant. Once we describe our adaptation algorithm for $(N_C,N_T)$
below, it will be clear how to adjust~\eqref{eq:p_est} to account for
the dynamic adaptation of $N_C.$

\subsection{Optimizing the duration of the contention and transmission phases}

In this section, we focus on the adaptation of $(N_C,N_T)$ based on
observed traffic statistics. This is to ensure the optimal balance
between the width of the contention phase and the transmission phase
so as to maximize the throughput of the system.

We assume that the frame duration $T$ is fixed, and the optimization
of $(N_C,N_T)$ is to be performed over a pre-defined set
$\mathcal{C},$ where $$\mathcal{C} \subseteq \{(N_C,N_T) \in \N \times
\N:\ N_C + k N_T = T\}.$$ Note that $\mathcal{C}$ is a finite set. Our
approach is to treat the optimization of $(N_C,N_T)$ over
$\mathcal{C}$ as a multi-armed bandit (MAB) problem, where the arms
correspond to the possible choices of $(N_C,N_T).$

Note that a MAB problem is characterized by a finite set of arms, each
arm $i$ being associated with an unknown reward distribution $F_i.$
Each time an arm $j$ is played, an independent reward drawn from $F_j$
is obtained. The goal is to choose which arm to play in each time
slot, seeking to maximize the aggregate reward obtained in the long
run. Of course, an oracle that knows the reward distributions of the
different arms would simply always play the arm with the highest mean
reward. However, since the reward distributions are unknown, MAB
algorithms have to estimate the mean reward of each arm by playing it
repeatedly. The MAB problem thus captures the classic trade-off between
exploration (i.e, playing each arm a large number of times to obtain
an accurate estimate of the mean reward), and exploitation (i.e.,
playing the arm that has so far provided the best mean reward). One
classical algorithm for the MAB problem is the Upper Confidence Bound
(UCB) algorithm \cite{burtini2015survey}. The regret associated with
this algorithm, which is defined as the (average) difference between
the aggregate reward obtained by always playing the `best' arm, and
the aggregate reward obtained by the algorithm, is known to be
$O(\log{N})$ over a horizon of $N$ plays. Note that the UCB plays any
suboptimal arm only $O(1/\log{N})$ fraction of the time (over $n$
plays). Moreover, the regret under UCB is near optimal --- it can be
shown that no online algorithm can have $o(\log{N})$ regret
\cite{lai1985asymptotically}.

We apply the UCB algorithm for $(N_C,N_T)$ adaptation as follows. The
set of arms is $\mathcal{C}.$ Each play of an arm corresponds to
operating the corresponding $(N_C,N_T)$ setting for $r$ successive
frames. The reward is proportional to the number of flows accepted
during the play. Specifically, we take the reward to be
$\frac{N_{acc}}{cTr},$ where $N_{acc}$ is the total number of accepted
flows over the $r$ frames in the play; note that the reward has been
normalized to lie in $[0,1].$ 

To ensure that the rewards are independent across plays, we complete
all flows that are active at the end of the $r$ frames in a short
sequence of \emph{flush frames}. No new flows are admitted during
these flush frames, which operate with $T/k$ transmission slots. Once
all active flows at the end of the play have been completed, the flush
frames stop and the next arm is selected.

The UCB algorithm operates as follows. In the beginning, all arms are
played once to create an initial estimate. In each subsequent round,
we pull the arm that has the highest estimated empirical reward up to
that point plus another term that is a decreasing function of the
number of times the arm has been played (for example, see
\cite{burtini2015survey}, \cite{auer2002finite}). Specifically, let
$m_{i,n}$ be the number of times arm $i$ has been played over $n$
plays. Let $r_n \in [0,1]$ be the reward we observe at play $n.$
Define $P_n\in \mathcal{C}$ be the choice of arm on the $n$th
play. The empirical reward estimate of arm $i$ after play $n$ is
\begin{equation*}
\hat{\mu}_{i,n} = \frac{\sum_{s=1}^{n} \indicator_{P_s = i} r_s}{m_{i,n}}.
\end{equation*}
UCB then assigns the following upper confidence bound value to each
arm $i$ at each time $n$:
\begin{equation*} 
%  \label{eq:ucb1}
\text{UCB}_{i,n} = \hat{\mu}_{i,n} + \sqrt{\frac{2\ln{n}}{m_{i,n}}}.
\end{equation*}
UCB algorithm selects, for play $n+1,$ the arm with the largest
upper confidence bound, i.e., 
\begin{equation*}
  P_{n+1} = \argmax\limits_{i \in \mathcal{C}} \text{UCB}_{i,n}.
\end{equation*}
%The second term in the upper confidence bound is inversely
%proportional to the number of times an arm is played. Thus, if the
%number of times an arm is played is large, $UCB_{i,t}$ value
%decreases. This enables the algorithm to continue exploring other
%arms.

Finally, we note that the contention probability adaptation described
earlier is specific to a particular arm. Thus, the $p$-adaptation is
performed independently for each arm, when it is played.

%\vspace{-2mm}
\section{Evaluation}
\label{sec:evaluation}

\begin{figure}[t]
	\begin{center}
	\includegraphics[width=\columnwidth]{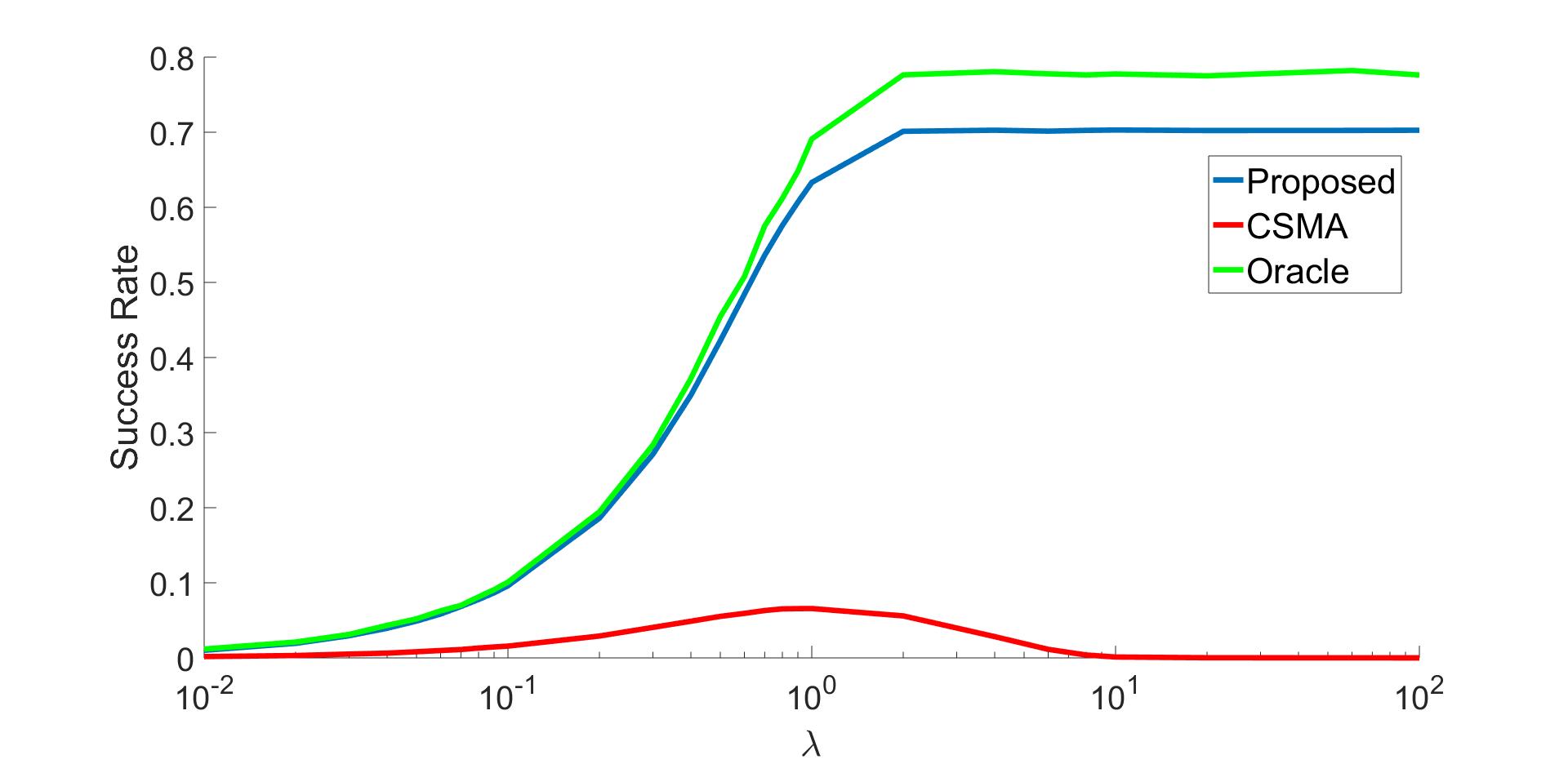}
	\caption{Deterministic load: Throughput v.s. $\lambda$}
	\label{fig:throughput_det}
	\end{center}
\end{figure}

\begin{figure}[t]
	\begin{center}
	\includegraphics[width=\columnwidth]{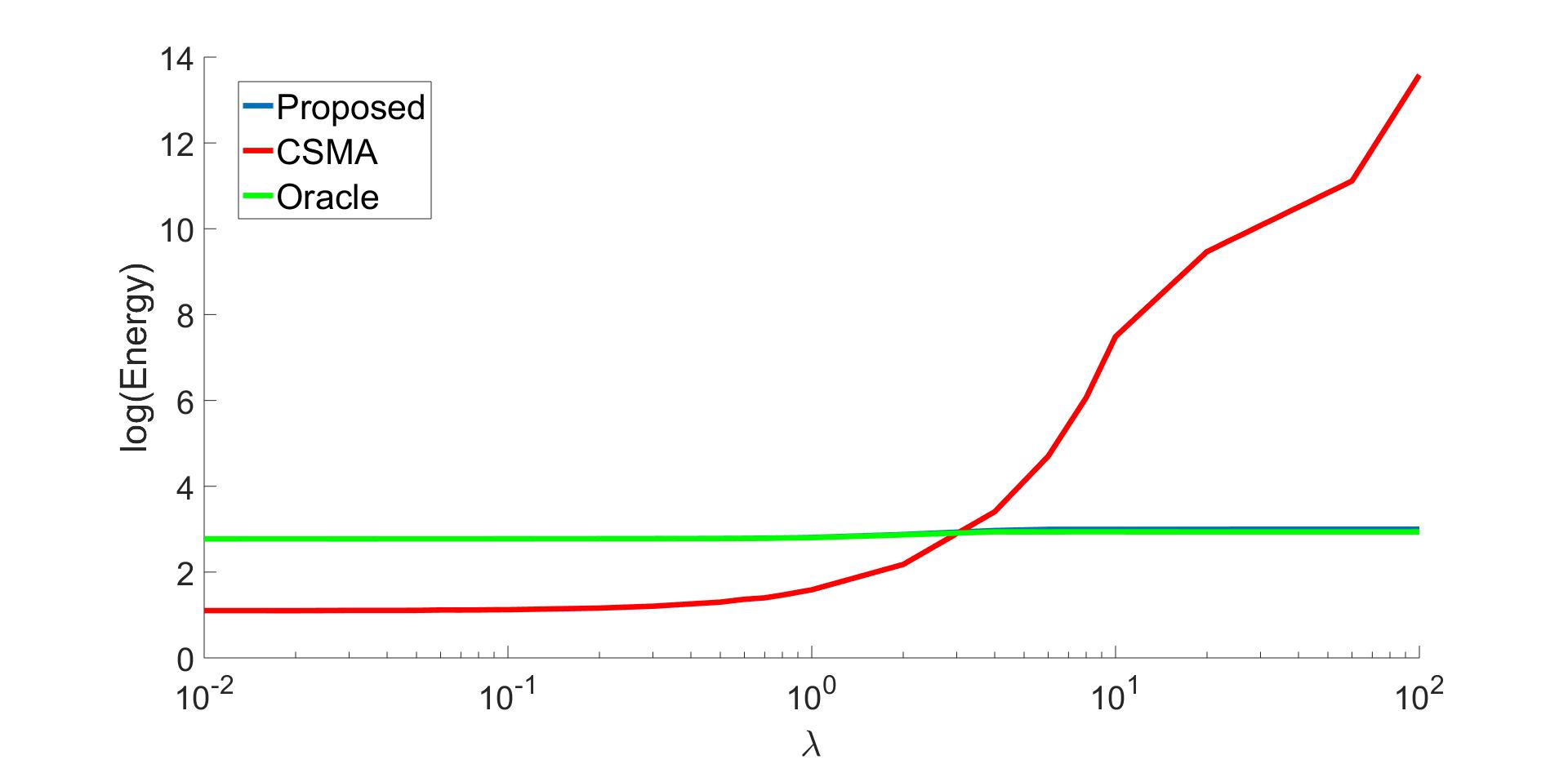}
	\caption{Deterministic load: Energy v.s. $\lambda$}
	\label{fig:energy_det}
	\end{center}
\end{figure}

\begin{figure}[t]
	\begin{center}
	\includegraphics[width=\columnwidth]{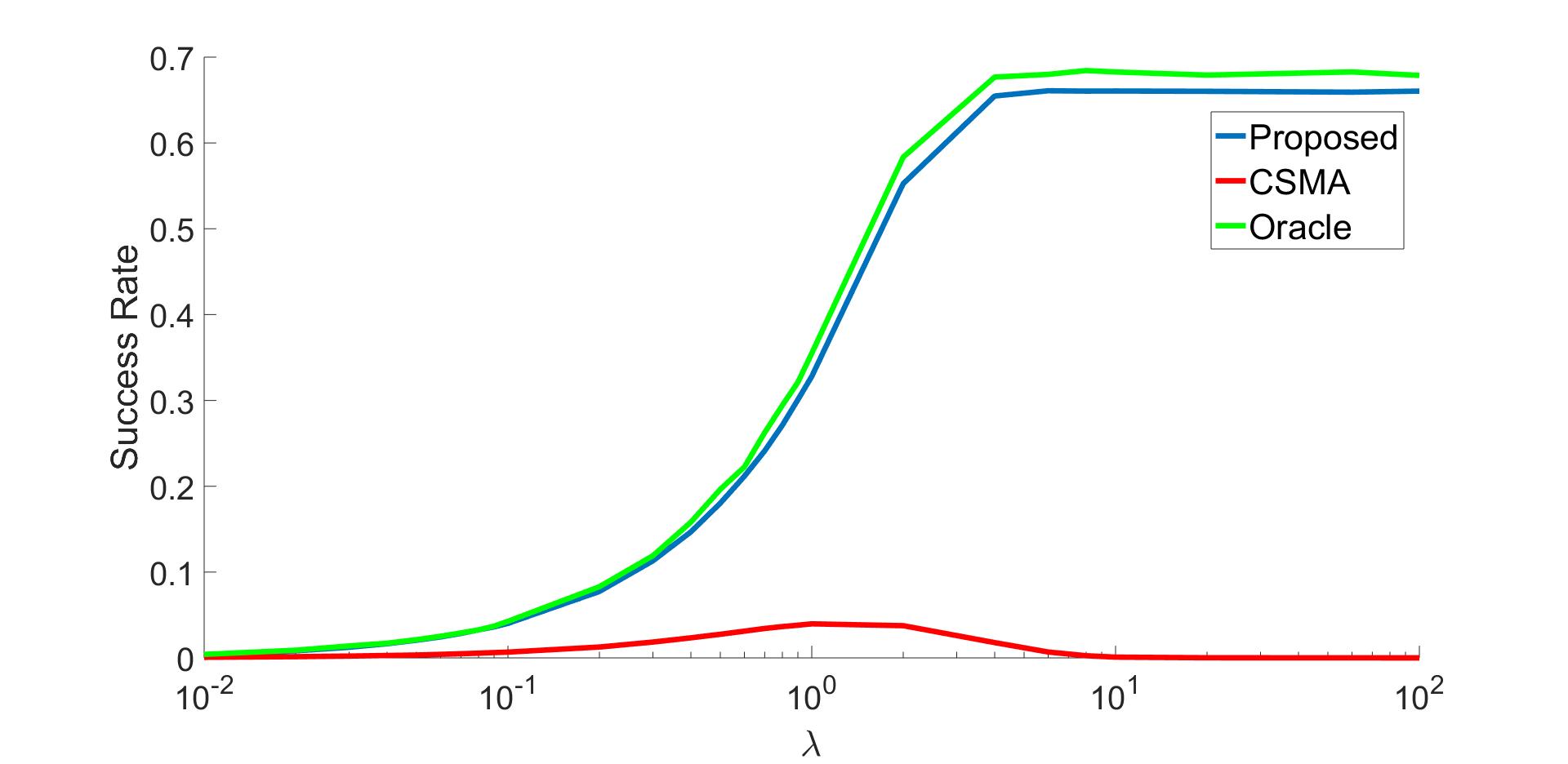}
	\caption{Geometric load: Throughput v.s. $\lambda$}
	\label{fig:throughput_geo}
	\end{center}
\end{figure}

\begin{figure}[t]
	\begin{center}
	\includegraphics[width=\columnwidth]{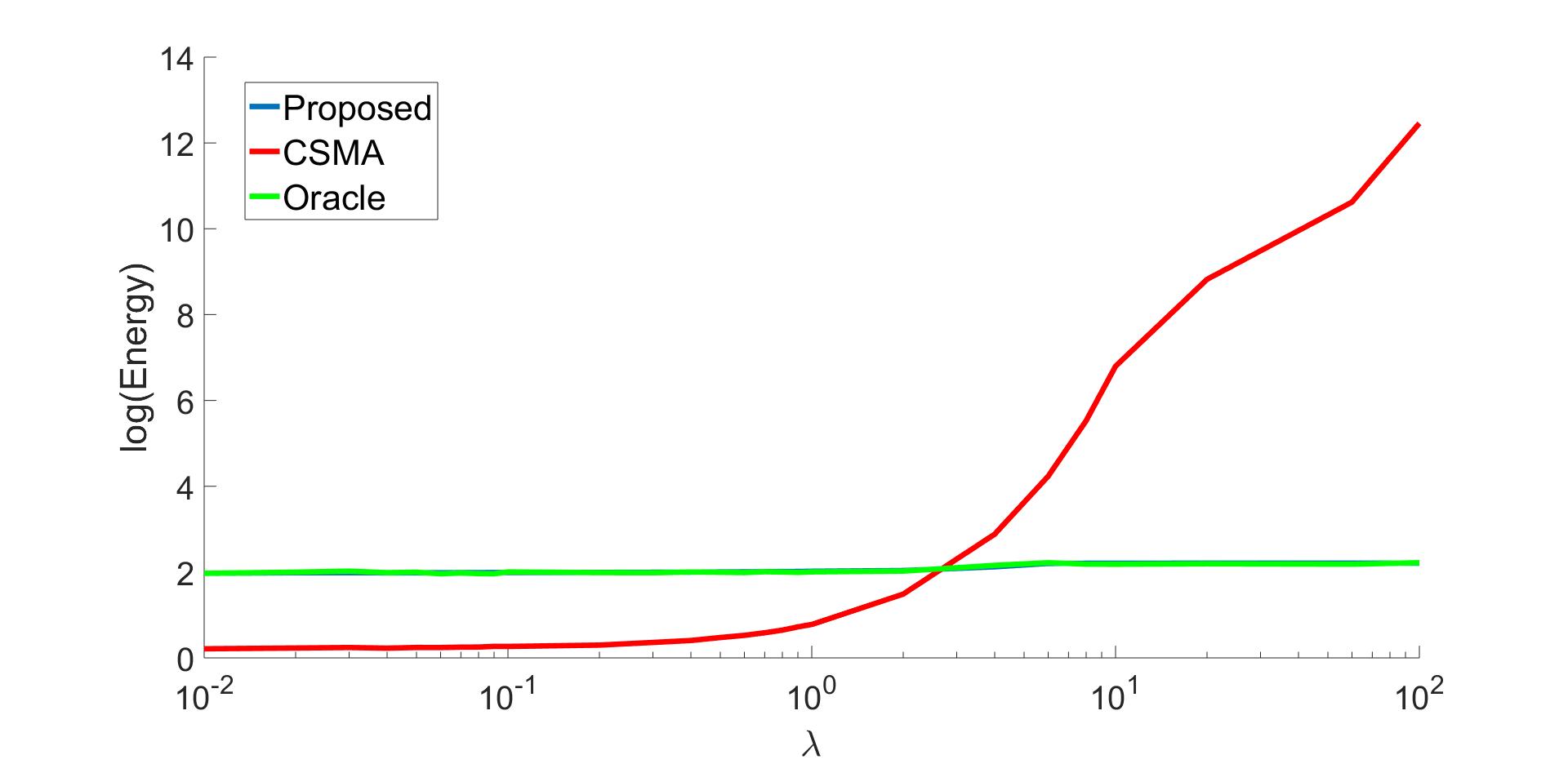}
	\caption{Geometric load: Energy v.s. $\lambda$}
	\label{fig:energy_geo}
	\end{center}
\end{figure}

\ignore{
\begin{figure*}
	\begin{center}
	  \resizebox{\linewidth}{!}{
          %			\subfloat[]{
          %		    	\includegraphics{images/throughput1.jpg}
          %            }
          %
           \subfloat[]{
	           \includegraphics{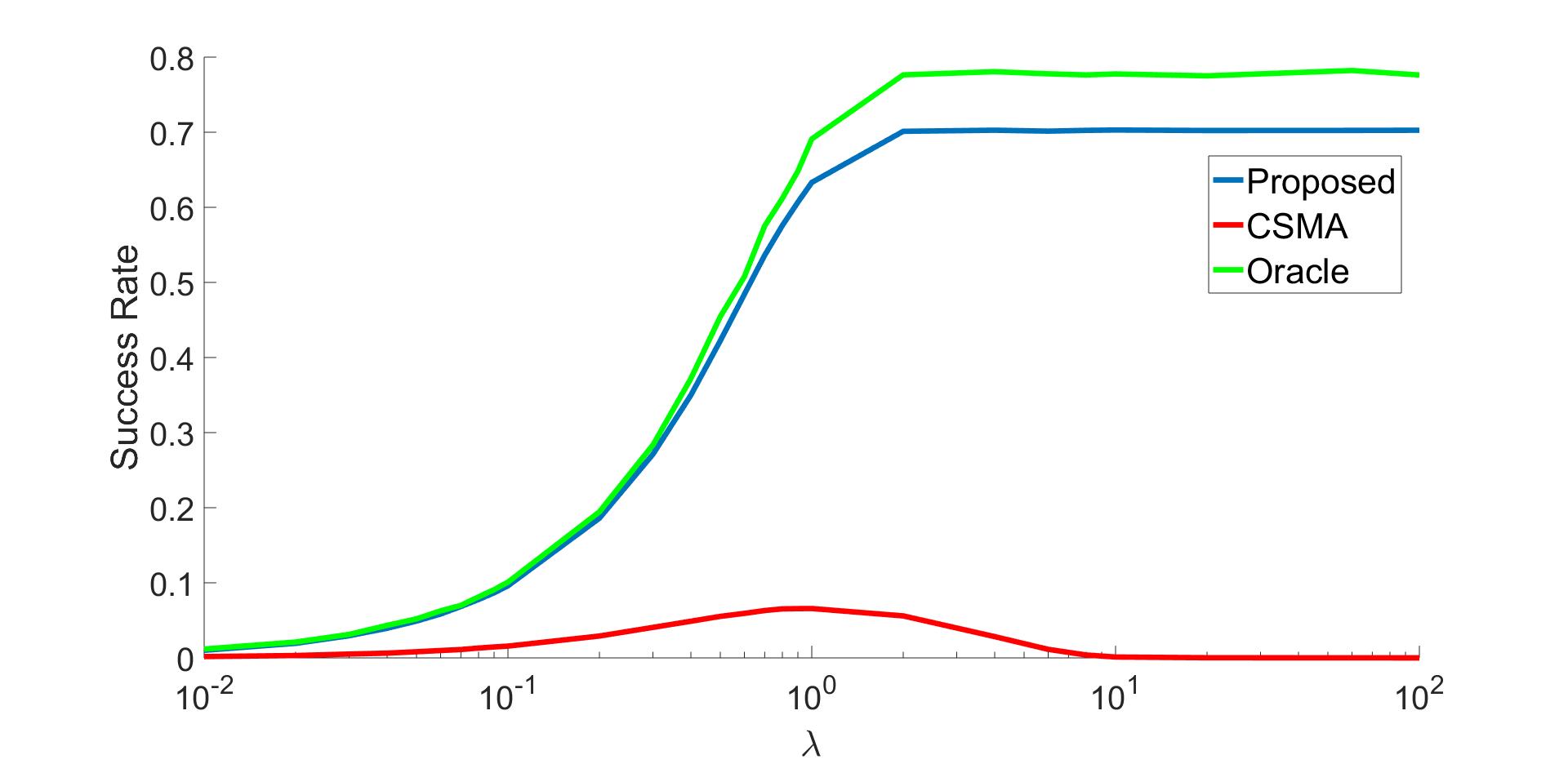}
           }
           \subfloat[]{
           	\includegraphics{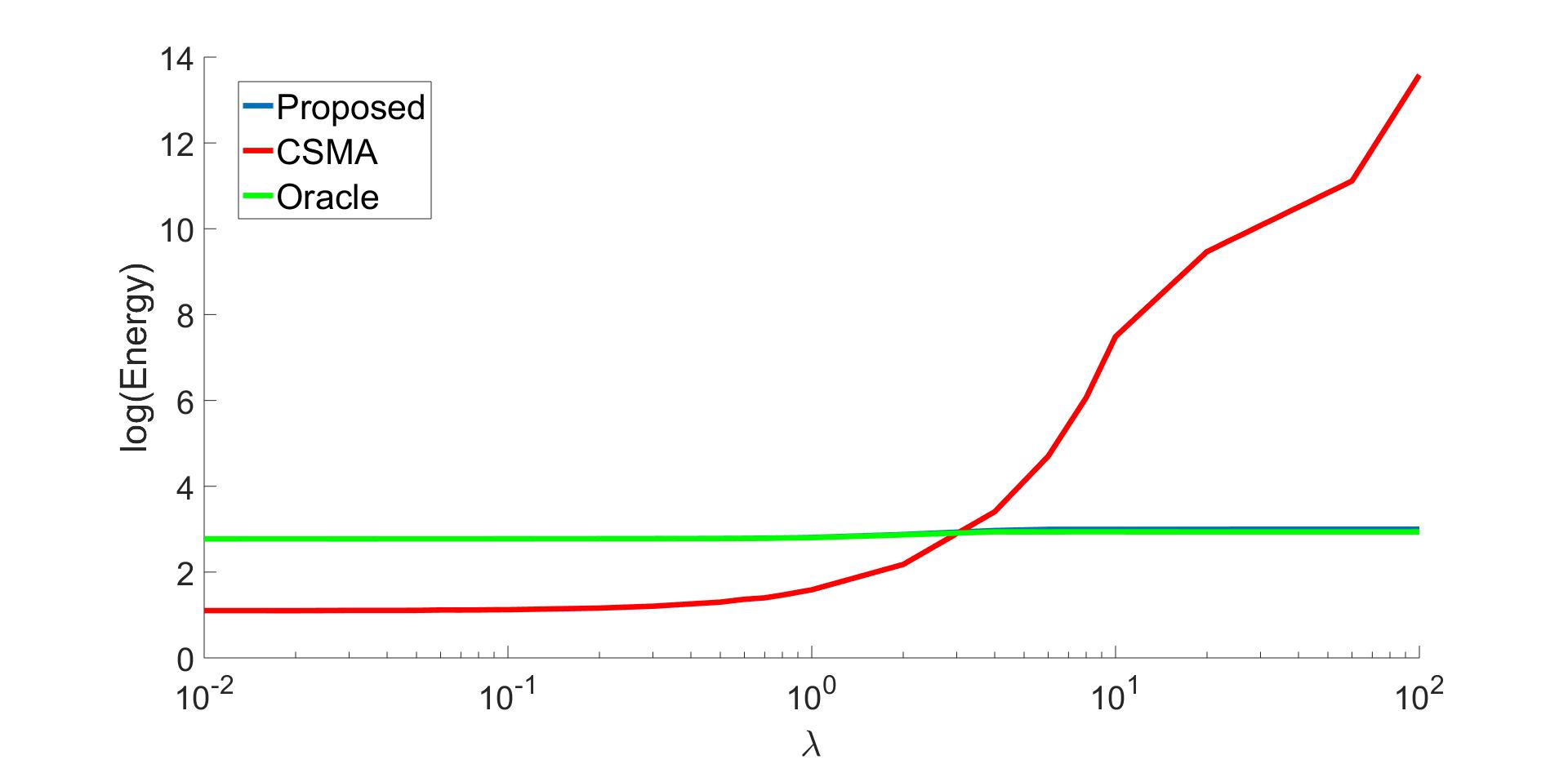}
           }
}
\caption{Determinstic load.}
		\label{fig:DetLoad}
	\end{center}
\end{figure*}

\begin{figure*}
	\begin{center}
		\resizebox{\linewidth}{!}{
                  %			\subfloat[]{
                  %		    	\includegraphics{images/throughput2.jpg}
                  %            }
            %
           \subfloat[]{
	           \includegraphics{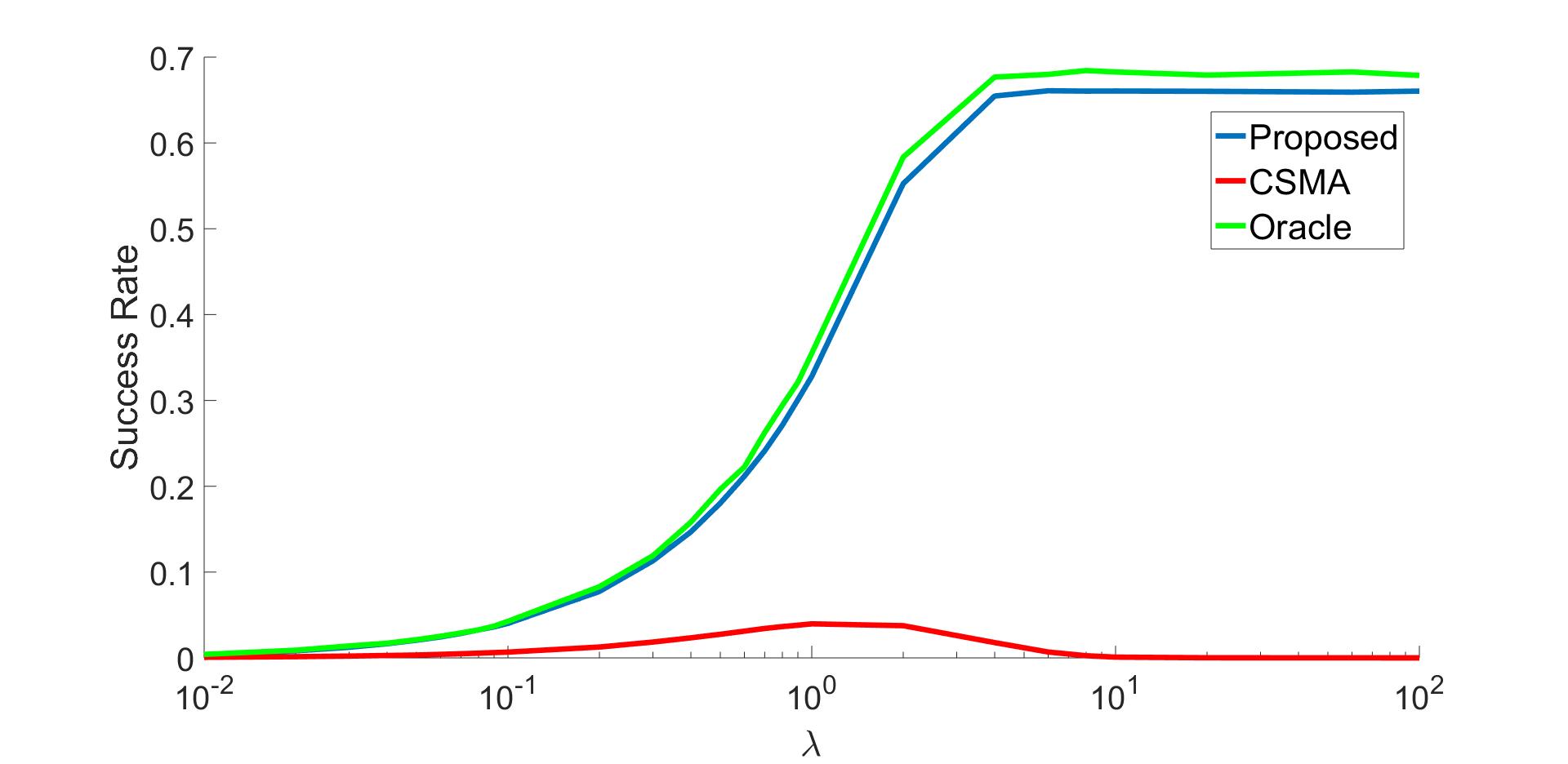}
           }
           \subfloat[]{
           	\includegraphics{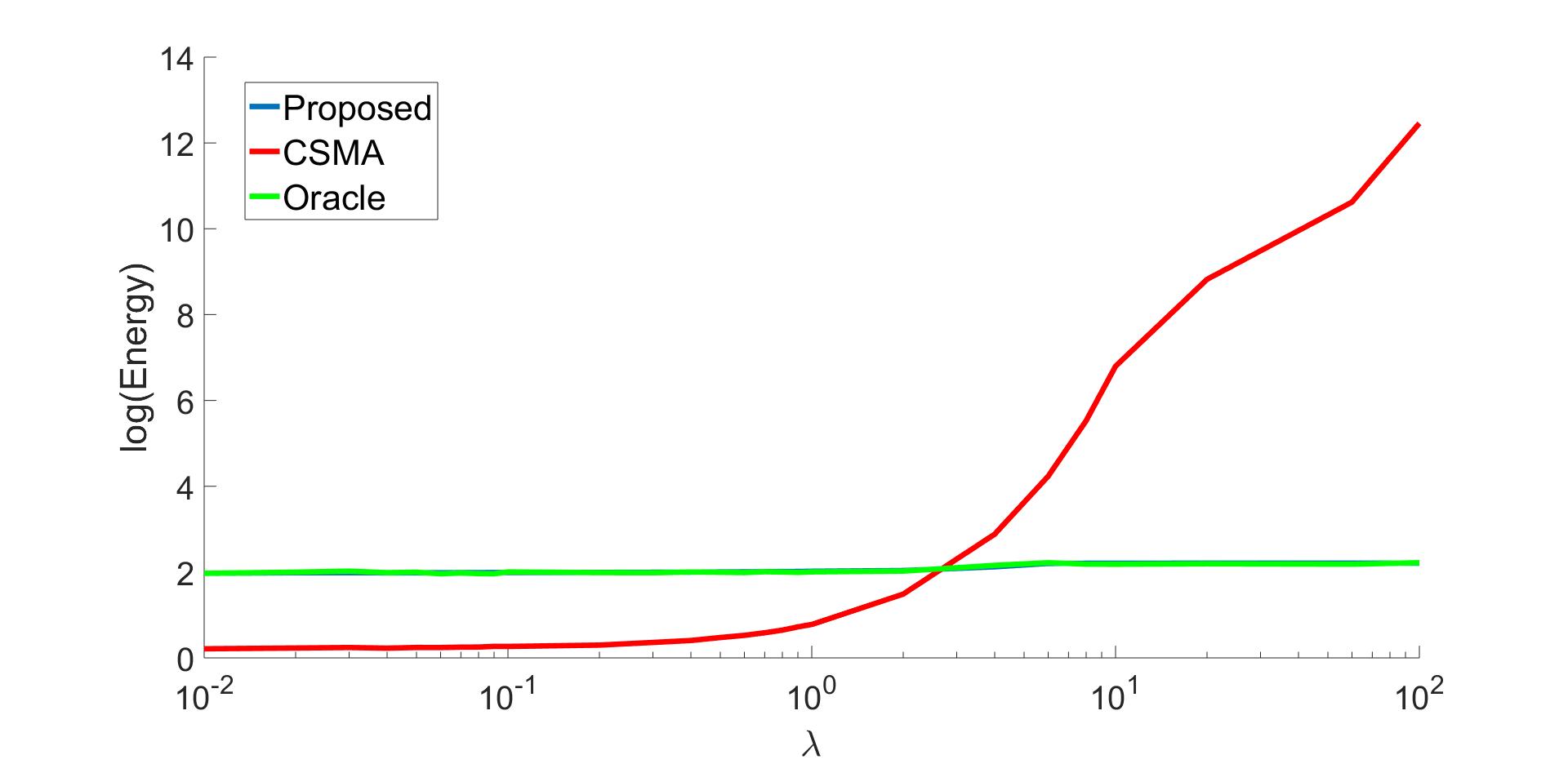}
           }
}
\caption{Geometrically distributed load}
		\label{fig:GeoLoad}
	\end{center}
\end{figure*}
}
The goal of this section is to evaluate the performance of the
protocol proposed in the previous sections alongside CSMA/CA via Monte
Carlo simulations.

The setting we used for these simulations is the following. We consider 3
channels, i.e., $c = 3.$ A single frame is taken to be 50 time units
long, with each transmission slot being five times the duration of a
contention slot. Thus, the number of transmission slots and contention
slots in a frame are constrained by $N_C + 5 N_T = 50.$ For parameter
adaptation, we consider the following possibilities for $(N_C,N_T):$
$$\mathcal{C} = \{(20,6), (15,7), (10,8),(5,9)\}.$$ Each $(N_C,N_T)$
configuration is run for $r = 50$ frames at a time (as part of a
single play of an arm under our MAB formulation).

For the comparison against CSMA/CA, we disregard channel hopping and
instead consider a single channel with a flow arrival rate of
$\lambda/c.$ (Equivalently, this may be viewed as CSMA/CA running in
parallel on $c$ channels, each experiencing $1/c$ fraction of the
traffic experienced by the proposed protocol.) We use the exponential
backoff model used in WiFi, with initial contention window set to
$CW_{\min} = 2$ and the maximum contention window set to $CW_{\max} =
16.$ Specifically, when a packet experiences a collision, it doubles
its contention window $CW$, and picks a uniformly distributed backoff
duration $b$ in $\{0,1,\cdots,CW-1\}.$ The next transmission is
attempted after the channel has been sensed idle for $b$ time units
(recall that one time unit is also the duration of a contention slot
in the proposed protocol). The duration of each packet transmission
under CSMA/CA is matched to the duration of a transmission slot under
the proposed protocol. If a packet suffers three successive
collisions, we abort the corresponding flow to maintain
stability. Moreover, once a flow misses its deadline under CSMA/CA, it
attempts no further transmissions (this limits the overhead on the
active flows in the system).
 
Finally, we evaluate the energy consumption of both protocols by
measuring the total transmission time across all flows over the
simulation horizon (including transmission during contention slots and
transmitted flows in the proposed protocol, and successful as well as
collision slots under CSMA/CA), normalized by the number of successful
flows. This yields a measure of the energy consumed per successful
flow by the system.

We consider two stochastic models for the flow parameters.

\noindent {\bf Scenario 1: Deterministic load}

Here, the load of each flow is deterministic and equal to~3. The slack
is taken to be uniformly distributed in the interval $[2, 20]$ (the
deadline is thus the deterministic load plus the randomly generated
slack). Figure~\ref{fig:throughput_det} depicts the variation of the
throughput of the system defined as the rate of successful flows
(i.e., number of successful flows over the simulation divided by the
simulation time) versus the arrival rate $\lambda$ for (i) the proposed protocol, (ii) an oracle variant of the proposed protocol that always operates the optimal $(p,N_C,N_T)$ values, (iii)
CSMA/CA. As expected, the throughput of the proposed protocol
saturates as the arrival rate grows, given the limited capacity of the
system. Moreover, the proposed scheme has a slightly lower saturation
throughput compared to the oracle version because of the imperfections
in the $p$ adaptation (note that we use a constant step-size), the
exploration cost of UCB, as well as the overhead involved in adapting
the $(N_C,N_T)$ values (the flush frames). In comparison, note that
the saturation throughput of CSMA/CA approaches zero as $\lambda$
increases. This is because as the rate of flow arrivals grows,
collisions become so prevalent that barely any flows succeed in
completing their transmissions before their deadline.

Figure~\ref{fig:energy_det} shows the energy consumption per
successful flow for all protocols as a function of $\lambda,$ on
a log-log scale. As we see, the energy efficiency of the proposed
protocol remains steady with increasing $\lambda,$ since our
reservation-based MAC only `wastes' energy during the contention
phase. Moreover, note that the energy efficiency closely matches the
oracle-based benchmark. In contrast, CSMA/CA has an energy per
successful flow that grows unboundedly with the arrival rate, due to
the steady energy consumption but dwindling throughput.

\noindent {\bf Scenario 2: Random load}

Next, we consider the case where the load of each flow is
geometrically distributed with a mean value of~1.25, with the slack
distribution unchanged. Figures~\ref{fig:throughput_geo}
and~\ref{fig:energy_geo} depict, respectively, the throughput and
energy (on a log scale) per successful flow for the same protocols. We see the same patterns as the deterministic load case.
%but note that the saturation throughput is slightly higher in this
%case. This is because in the present setting, owing to the
%variability of the load, the proposed protocol is able to admit more
%flows by admitting those with a smaller load.

We conclude this section by noting that the proposed protocol
outperforms CSMA/CA with respect to throughput (equivalently,
QoS-compliance) as well as energy efficiency. This is particularly
true in heavy traffic, where the number of nodes attempting to access
the channel at any time is large (different from the settings in which
WiFi is presently deployed), when the overhead associated with the
completely distributed scheduling by CSMA/CA becomes prohibitive.

%\vspace{-2mm}
\section{Concluding remarks}
%\vspace{-2mm}

In this paper, we propose a reservation-based MAC framework for
IoT/M2M applications, where flows are scheduled centrally in a
deadline-aware manner by a central master node, and MAC parameters
are adapted dynamically based on the statistics of the observed
traffic. We demonstrate that such a MAC outperforms CSMA/CA when the
number of connected devices becomes large, as is projected in the IoT
regime.

Note that the proposed framework can co-exist alongside conventional
WiFi --- WiFi being used to connect a few user-operated
(high-bandwidth) devices, and the proposed reservation-based MAC being
used to connect the (large number of) IoT devices.

Future work will focus on (i) proving the throughput optimality of the
proposed schemes, (ii) extending to flows that are periodic in nature, and
(iii) generalizing to the case of multiple interfering networks, each
with their own master node, where the masters dynamically allocate
channels among themselves based on their observed congestion as well
as interference constraints.

%\input{appendix}

%{\small
\bibliographystyle{IEEEtran}
\bibliography{refs,PSrefs}
%}
\end{document}